\documentclass[aps,twocolumn]{revtex4}
\usepackage{epsf}
 
\begin{document}

\title{Spectroscopic Bogoliubov features near the unitary limit}   

\author{T.\ Doma\'nski}
\affiliation{
       Institute of Physics, M.\ Curie-Sk\l odowska University, 
       20-031 Lublin, Poland}

\date{\today}

\begin{abstract}
We analyze the single particle excitation spectrum of the ultracold 
fermion atom system close to the unitary limit where there has been 
found experimental evidence for the Bogoliubov quasiparticles below 
as well as above the transition temperature $T_{c}$. We consider 
the short-range correlations originating from the preformed pairs 
and try to reproduce the experimental data adapting phenomenological 
selfenergy previously used for description of the anti-nodal spectra 
of the underdoped cuprate superconductors. We show that it 
fairly fits the lineshapes obtained by 
the momentum resolved RF spectroscopy for $^{40}$K atoms.
\end{abstract}

\maketitle

\section{Introduction} 

Spectroscopic tools such as the Bragg scattering technique 
\cite{Bragg}, the RF-pulse spectrometry \cite{Chin-04} and 
its $k$-resolved improvement \cite{Stewart-08} were able to 
provide a clear-cut evidence for the superfluid nature of
the ultracold fermion atom systems. Especially intriguing 
among the obtained data is the quasiparticle back-bending 
dispersion near the Fermi momentum observed below and above 
the superfluid transition temperature $T_{c}$  \cite{Jin-10}. 
This fact indicates that the Bogoliubov-type quasiparticles 
survive even in a normal state where the long-range coherence 
between fermion pairs does no longer exist. 

Similar fingerprints of the dispersive Bogoliubov quasiparticles 
have been previously detected above $T_{c}$ also in the underdoped 
cuprate superconductors by the measurements of ARPES \cite{ARPES-08} 
and the Fourier-transformed STM  \cite{Davis-09}. They confirmed 
expectations motivated by the Uemura scaling $T_c \propto n_s$ 
\cite{Uemura} and later on supported by the residual Meissner 
rigidity seen above $T_{c}$ in the tera-Hertz \cite{Orenstein-99}  
and the torque magnetometry \cite{torque} experiments.  
Superconducting transition of the underdoped cuprates seems
hence to be controlled not by the pair-formation but rather 
by onset of the phase coherence. This point is however still 
a controversial issue. 

In the present work we consider the spectroscopic Bogoliubov 
features common above $T_{c}$ for the ultracold fermion gases 
and underdoped cuprate materials taking into account the 
short-range correlations driven by preformed fermion pairs. 
Such problem is currently widely discussed in the literature 
\cite{Chen-09,Pieri-09,Schneider-10,Ohashi-10,Haussmann-09} 
(for a comprehensive discussion see e.g.\ \cite{Levin-10} 
and other references cited therein). We shall present the 
results obtained for the single particle excitations using 
the selfenergy motivated by the local solution of the Feshbach
coupling and also suggested by perturbative studies 
of the pairing fluctuations 
\cite{Levin-10,pairing-ansatz,Senthil-09}.

We start with analysis of the exact solution for the local 
Feshbach scattering problem.  We next discuss how this 
result can be cast on the itinerant case. 
% adopting the ansatz used in analogous treatments 
% of the pairing fluctuations \cite{Senthil-09}. 
Introducing the phenomenological scattering 
rate we then try to reproduce the single particle spectra 
for temperatures corresponding to the experiment of 
the Boulder group \cite{Jin-10}. Summarizing our results 
we point out the problems relevant for future studies.

\section{Local scattering on pairs} 

The momentum-resolved spectroscopic measurements of the Boulder 
group  \cite{Jin-10} have been done using $^{40}$K atoms equally 
populated in the hyperfine states $\left| 9/2,-9/2 \right>$ and 
$\left| 9/2,-7/2\right>$ (we shall denote them symbolically as 
$\sigma=\uparrow$ and $\sigma=\downarrow$). By applying magnetic 
field the atoms were adiabatically brought to vicinity of the
unitary limit, slightly on the BEC side $\left( k_F a \right)
^{-1} =0.15$. Under such conditions energies of the atoms 
are nearly degenerate with the weakly bound molecular configurations. 
The single atoms and molecules are there strongly scattered 
from each other through the conversion processes.

At a given position ${\bf r}$ in the magneto-optical trap 
such Feshbach resonant interactions can be described by 
the following local Hamiltonian \cite{Chiofalo-02}
\begin{eqnarray}
\hat{H}_{loc}({\bf r}) & = & \sum_{\sigma =\uparrow, \downarrow} 
\varepsilon({\bf r}) \; \hat{c}^{\dagger}_{\sigma}({\bf r}) 
\hat{c}_{\sigma}({\bf r}) + E({\bf r}) \; \hat{b}^{\dagger}({\bf r})  
\hat{b}({\bf r}) \nonumber \\ &+& g 
\left(  \hat{b}^{\dagger}({\bf r}) \hat{c}_{\downarrow}({\bf r}) 
\hat{c}_{\uparrow}({\bf r}) +  \hat{c}^{\dagger}_{\uparrow}
({\bf r}) \hat{c}^{\dagger}_{\downarrow}({\bf r}) \hat{b}({\bf r})
\right)
\label{Andreev}
\end{eqnarray}
where $g$ denotes the $s$-wave channel scattering strength,
$\hat{c}_{\sigma}^{(\dagger)}({\bf r})$ are fermion operators of 
the single atoms in two hyperfine states $\sigma$=$\uparrow$, $\downarrow$ 
and operators $\hat{b}^{(\dagger)}({\bf r})$ correspond to the molecular 
state. Spatial variation of the energies $\varepsilon({\bf r})$, 
$E({\bf r})$ comes from the trapping potential and usually 
take the parabolic dependence with some characteristic radial 
and axial frequencies.

Hilbert space of the local Hamiltonian (\ref{Andreev}) 
is spanned by four fermion configurations $\left| F \right>
\!=\!\left| 0 \right>$, $\left| \uparrow \right>$, $\left| 
\downarrow \right>$, $\left| \uparrow \downarrow \right>$ 
and two molecular ones $\left| B \right)\!=\!\left| 0 \right)$, 
$\left| 1 \right)$ - altogether 8 states. Six of these
states $\left| F \right> \otimes \left| B \right)$ are
eigenfunctions of (\ref{Andreev}) and two vectors $\left| 
\uparrow \downarrow \right> \otimes \left| 0 \right)$ 
and $\left| 0 \right> \otimes \left| 1 \right)$ get mixed
by the Feshbach interaction. With the suitable unitary 
transformation we can determine the true eigenfunctions 
\begin{eqnarray}
\left|  \psi_A \right> & = &
u \left| 0 \right> \otimes \left| 1 \right) 
+ v \left| \uparrow \downarrow \right> \otimes \left| 0 \right) 
\label{A} \\
\left| \psi_B \right> & = &
-v \left| 0 \right> \otimes \left| 1 \right)  
+ u \left| \uparrow \downarrow \right> \otimes \left| 0 \right) 
\label{B}
\end{eqnarray}
where $u^2,v^2= \frac{1}{2} \left[1 \!\pm\! (\varepsilon \!-\! E/2)
/\sqrt{(\varepsilon \!-\! E/2)^2+g^2}\right]$ and the eigenvalues 
are given by $\varepsilon \!-\! E/2 \pm \sqrt{(\varepsilon \!-\! E/2)
^2+g^2}$. The equations (\ref{A},\ref{B}) are reminiscent of 
the Bogoliubov - Valatin transformation of the standard BCS 
problem, where true quasiparticles are represented by linear 
combinations of the particle and hole states. In our present 
case the  additional degree of freedom related to the molecular 
state causes qualitative differences discussed below.

\begin{figure}
\hspace{-1cm}{\epsfxsize=8cm\centerline{\epsffile{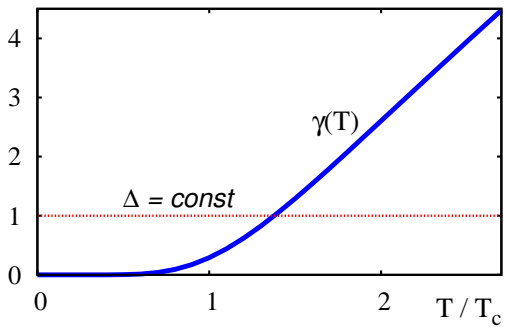}}}
\caption{(Color online) Temperature dependence of the 
pheno\-menological scattering rate $\gamma(T)$ introduced 
in equation (\ref{gamma_T}) and closely resembling the proposal 
of the Ref.\ 
\cite{Senthil-09}.}
\label{gammas}
\end{figure}

Using the spectral Lehmann representation we can exactly 
determine  the single particle Green's function ${\cal{G}}_{loc}(\tau)
=-\langle T_{\tau} \hat{c}_{\sigma}({\bf r},\tau) \hat{c}^{\dagger}
_{\sigma}({\bf r},0) \rangle$. Its Fourier transform takes
the three-pole structure 
\begin{eqnarray}
{\cal{G}}_{loc}(i\omega_{n})& = & \left[ 1\!-\!Z(T)\right] 
\left( \frac{u^{2}}{i\omega_{n}\!-\!\varepsilon_{+}} +  
\frac{v^{2}}{i\omega_{n}\!-\!\varepsilon_{-}} \right)
\nonumber \\
&+ & \frac{Z(T)}{i\omega_{n} \!-\! \varepsilon} ,
\label{local}
\end{eqnarray}
where $\varepsilon_{\pm}=E/2\pm \sqrt{(\varepsilon \!-\! 
E/2)^2+g^2}$ and explicit form of $Z(T)$ was given by us in Ref.\ 
\cite{local-solution}. Let us now focus on $E=0$, i.e.\ 
the unitary limit case. The single particle spectral function 
$- \frac{1}{\pi} \mbox{\rm Im} \left\{ {\cal{G}}_{loc}
(\omega+i0^{+})\right\}$ consists then of: 
\begin{itemize}
\item[{a)}] a remnant of the free particle state at $\omega\!=\!
  \varepsilon$ with the temperature dependent residue $Z(T)$, and
\item[{b)}] Bogoliubov-type quasiparticles at  $\omega\!=\!
  \pm\sqrt{\varepsilon^2 +g^2}$ whose spectral weights are
  $\left[1\!-\!Z(T)\right]u^{2}$ and  correspondingly  
  $\left[1\!-\!Z(T)\right]v^{2}$.
\end{itemize}  
The free particle residue $Z(T)$ is sensitive to temperature. 
For instance, at $\varepsilon \!=\!0$ we have $Z(T)=\frac{2}
{3+\cosh{\left(g/k_{B}T\right)}}$ \cite{local-solution} which 
vanishes exponentially when $T \! \rightarrow \! 0$. 
It means that at low temperatures only the Bogoliubov-type 
quasiparticles are present. Upon increasing temperature 
an amount $Z(T)$ of the spectral weight is transferred from
the Bogoliubov quasiparticles to the free fermion state,
effectively filling-in the gaped spectrum.

\section{Similarity to other studies}

Our exact solution of the local Feshbach scattering problem
(\ref{Andreev}) coincides with physical conclusions obtained 
by T.\ Senthil and P.A.\ Lee \cite{Senthil-09} who have explored 
influence of the incoherent pairs (preformed above $T_c$) on 
the single particle spectrum. The local pair operator 
$\hat{F}({\bf r},t)\equiv \hat{c}_{\downarrow}({\bf r},t)
\hat{c}_{\uparrow}({\bf r},t)$  can be formally represented 
through the amplitude and phase
\begin{eqnarray}
\hat{F}({\bf r},t)
= \hat{\chi}({\bf r},t) \; e^{i\hat{\phi}({\bf r},t)} .
\label{ilp}
\end{eqnarray}
Since above $T_c$ the pairs need not be dissociated $\chi \neq 0$ 
their phase $\phi({\bf r},t)$ must be randomly oriented, precluding 
any off-diagonal long-range order (ODLRO) $\langle \hat{F}
({\bf r},t) \rangle \!=\!0$. To account for superconducting 
fluctuations the authors assumed  certain temporal 
$\tau_{\phi}$ and spacial $\xi_{\phi}$ scales, over which 
the pairs are short-range correlated
\begin{eqnarray}
\langle  
\hat{F}^{\dagger}({\bf r},t)
\hat{F}({\bf 0},0)
%\hat{c}_{\uparrow}^{\dagger} ({\bf r},t)
%\hat{c}_{\downarrow}^{\dagger}({\bf r},t)
%\hat{c}_{\downarrow} ({\bf 0},0)
%\hat{c}_{\uparrow}({\bf 0},0) 
\rangle \propto |\chi|^2 
\mbox{\rm exp}\left( - \frac{|t|}{\tau_{\phi}} 
- \frac{|{\bf r}|}{\xi_{\phi}} \right) .
\label{decay}
\end{eqnarray}
Taking into account the pairing field (\ref{decay}) by means
of the lowest order perturbative scheme they have determined 
the single particle Green's function ${\cal{G}}({\bf k},
i\omega_{n})=\left[ i\omega_n - \varepsilon_{\bf k} - 
\Sigma({\bf k},i\omega_{n})\right]^{-1}$ interpolating it 
by \cite{Senthil-09}
\begin{eqnarray}
\Sigma({\bf k},i\omega_{n})= - \Delta^2 \frac{i\omega_n - 
\varepsilon_{\bf k}}{ {\omega_{n}^{2}+
\varepsilon_{\bf k}^{2}+\pi\Gamma^2}} \;\; ,
\label{interpol}
\end{eqnarray}
%
%given by equation (16) in the Ref.\ \cite{Senthil-09}. 
where $\Delta \propto |\chi|$ is a magnitude of the energy 
gap due to pairing and parameter $\Gamma$ is related to the 
in-gap states. At low energies (i.e.\ for $|\omega| \ll 
\Delta$) a dominant contribution of the spectrum comes 
from the in-gap quasiparticle with  residue 
$Z\!\equiv\!\left( 1+ \frac{\Delta^{2}}{\pi \Gamma^2} 
\right)^{-1}$ whereas at higher energies the BCS-type 
quasiparticles are formed. 
% Partial reduction of their spectral weight 
% has to obey the sum rule. 
All these features are present in the exact solution 
(\ref{local}) of the local Feshbach scattering problem 
(\ref{Andreev}) for which we have
\begin{eqnarray}
\Sigma_{loc}(i\omega_n) = - \left[1\!-\!Z(T)\right]
g^{2} \frac{i\omega_n - \varepsilon}{ {\omega_{n}^{2}
+\varepsilon^{2}+Z(T)g^2}} \;\; .
\label{sigma_loc}
\end{eqnarray}
%
%We can hence regard $g$  as a measure of the gap energy 
%$\Delta$.

\begin{figure}
\epsfxsize=12.5cm\centerline{\epsffile{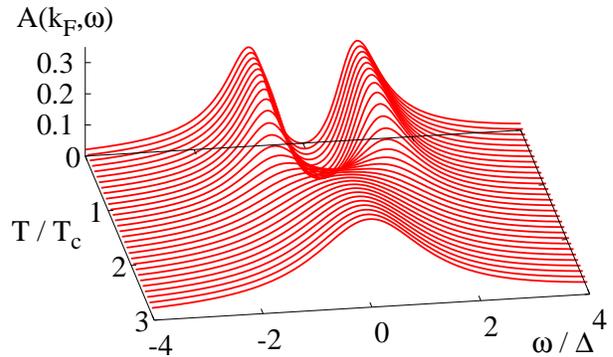}}
\caption{(Color online) Evolution of the spectral function 
$A({\bf k},\omega)$ at the Fermi momentum ${\bf k}\!=\!{\bf k}_{F}$
for temperature regime $0 \! \leq \! T \! \leq \! 3T_{c}$. 
The gaped superconducting spectrum smoothly  evolves into 
the single peak structure nearby $1.5T_{c}$.}
\label{kF}
\end{figure}

\begin{widetext}
\begin{figure}
\parbox{19cm}{
{\epsfxsize=11cm {\epsffile{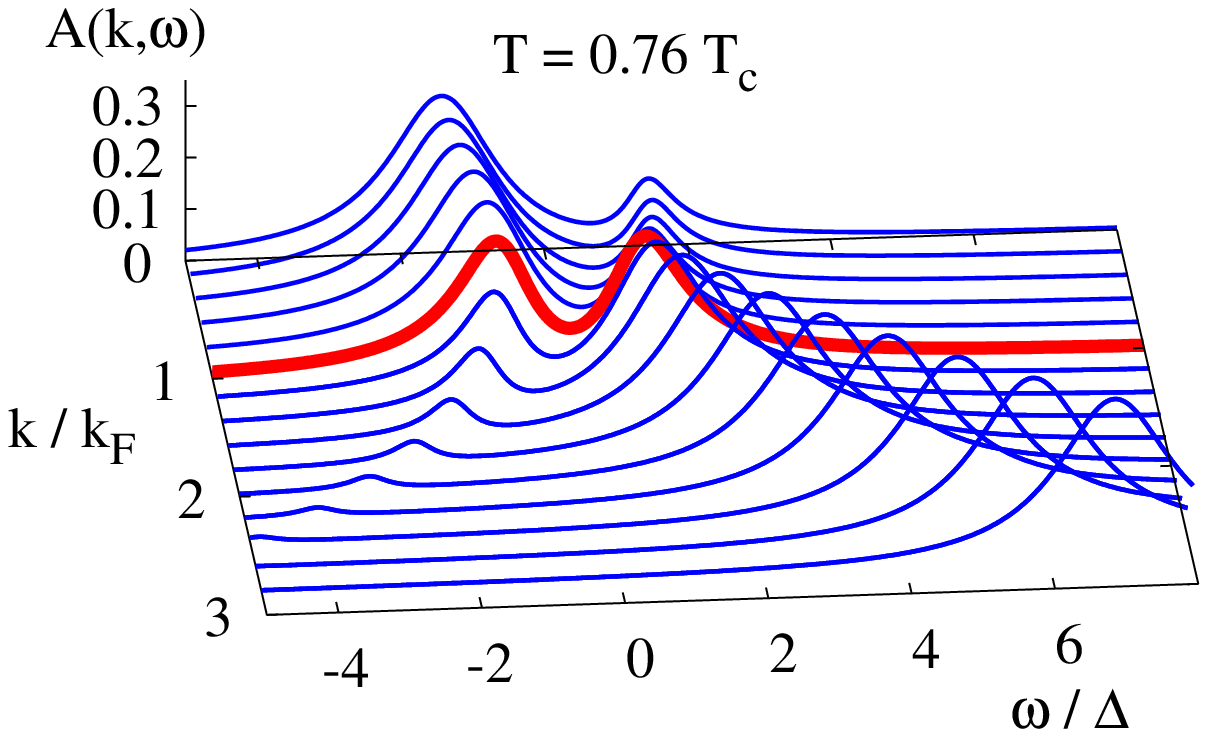}}}\hspace{-3.5cm}
{\epsfxsize=11cm {\epsffile{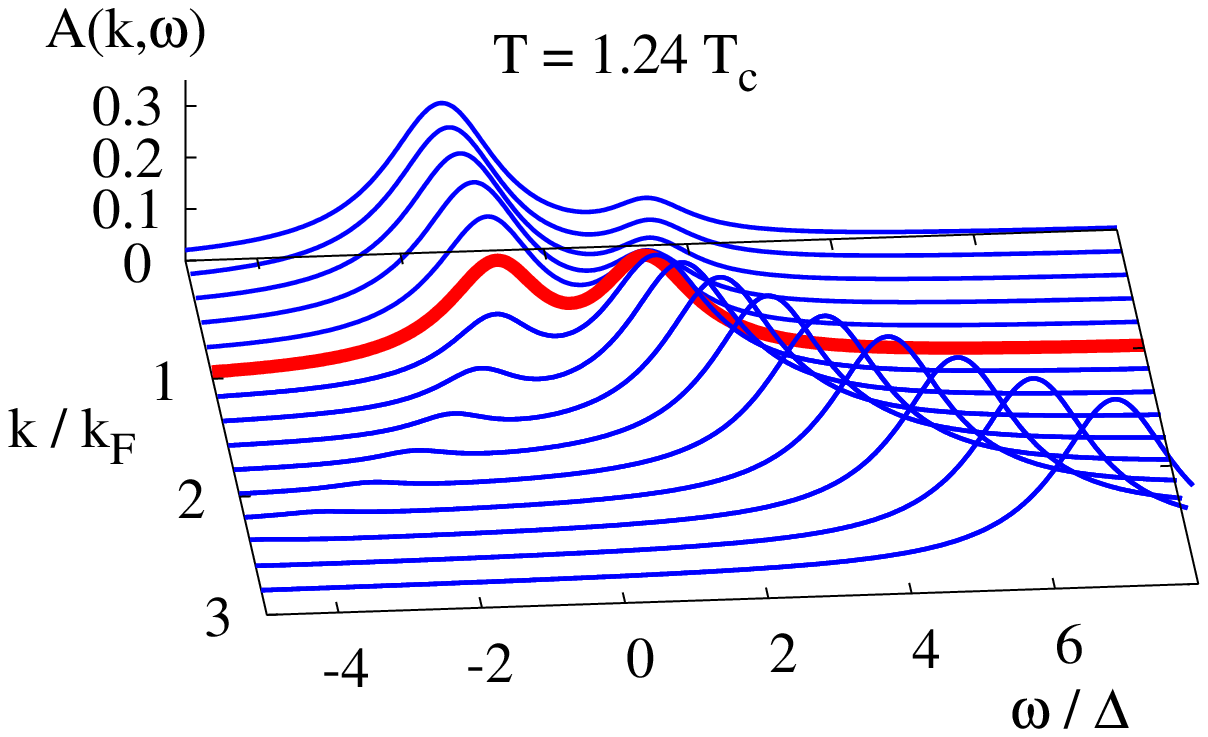}}}\\
{\epsfxsize=11cm {\epsffile{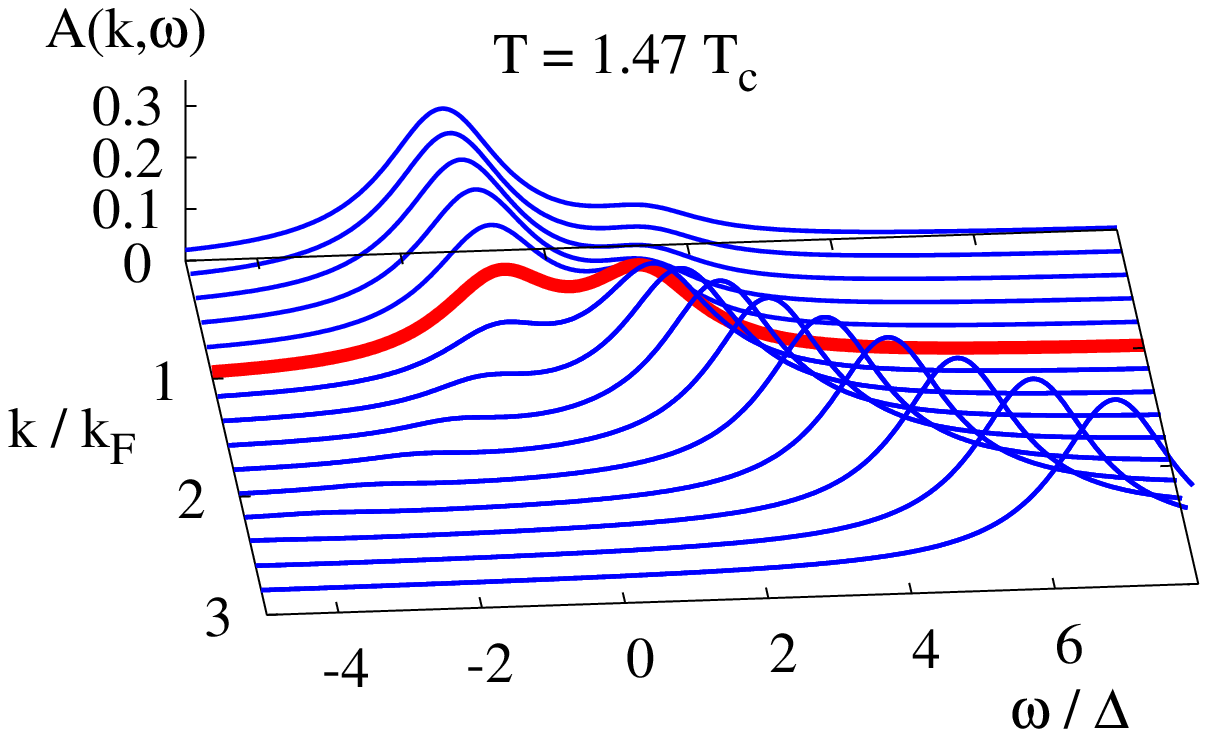}}}\hspace{-3.5cm}
{\epsfxsize=11cm {\epsffile{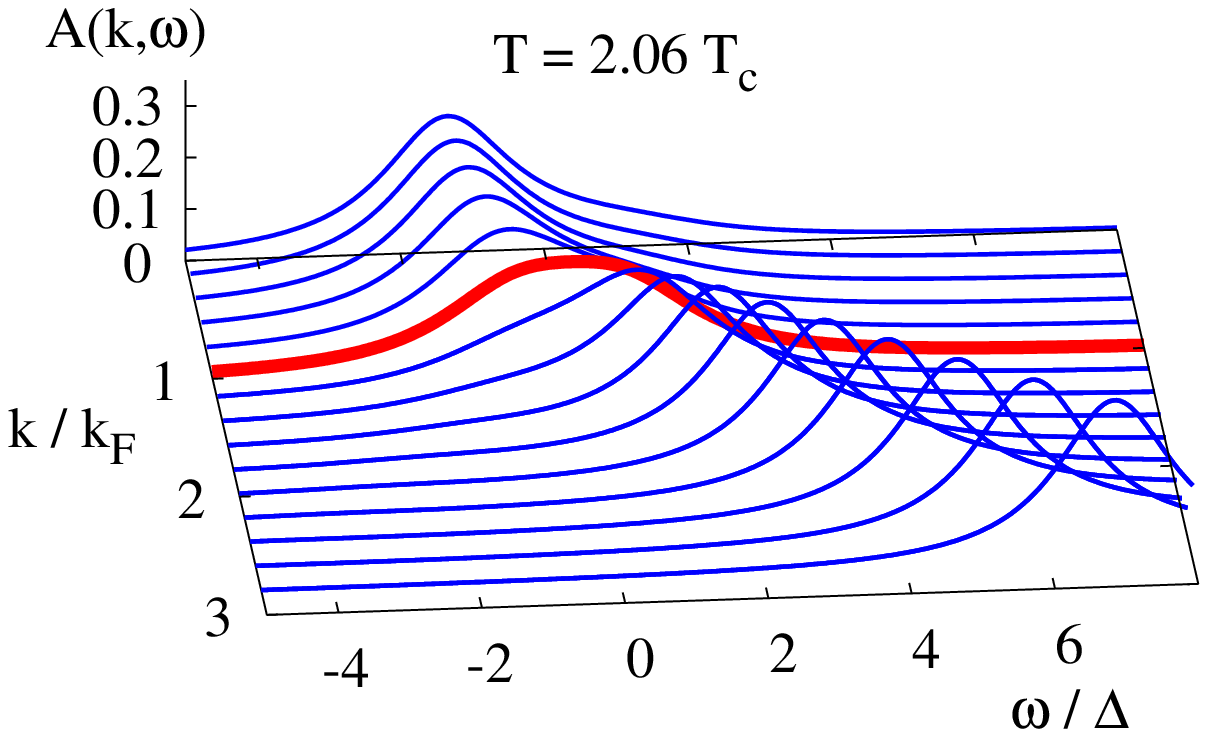}}}}
\caption{(Color online) Momentum and energy dependence 
of the spectral function $A({\bf k},\omega)$ for the set 
of temperatures reported experimentally by the Boulder group 
\cite{Jin-10}. We can notice that the Bogoliubov quasiparticle 
features (bending-down dispersion) is preserved to nearly $1.5T_c$.}
\label{spectral_fun}
\end{figure}
\end{widetext}

\section{Phenomenological pairing ansatz}

Combining the local physics (\ref{Andreev}) with the itinerancy 
$\hat{T}_{kin}({\bf r})$ of fermions and molecules is a highly 
non-trivial task. Certain aspects of the complete Hamiltonian 
\begin{eqnarray}
 \hat{H}= \int d{\bf r} \left( \hat{T}_{kin}({\bf  r}) 
 + \hat{H}_{loc}({\bf  r}) \right) 
\label{Hamiltonian}
\end{eqnarray}
have been so far addressed by:
the selfconsistent perturbative treatment \cite{Robin}, 
dynamical mean field theory \cite{Romano}, 
self-consistent T-matrix approach \cite{Micnas},  
conserving diagrammatic approximations \cite{Levin-10}
RG-like scheme \cite{Domanski}, 
path integral formulation for the bond operators \cite{Cuoco-06} 
and several other techniques. Some of these studies directly
\cite{Domanski,Cuoco-06} or indirectly \cite{Levin-10}
pointed at the Bogoliubov quasiparticles surviving 
above $T_{c}$.

Here we would like to focus on the main physical outcomes
which could be relevant to the experimental situation of 
the Boulder group \cite{Jin-10}. For this purpose we apply 
the phenomenological selfenergy
\begin{eqnarray}
\Sigma({\bf k}, \omega)=\frac{\Delta^2}
{\omega+\varepsilon_{\bf k}+i\;\gamma(T)} - i\;\Sigma_0
\label{ansatz}
\end{eqnarray}
which, according to the argumentation outlined in section 
III of the Ref.\ \cite{Senthil-09}, originates from 
(\ref{interpol}) and similarly (\ref{sigma_loc}). 
The particular structure (\ref{ansatz}) has been also 
suggested by previous studies of the precursor pairing 
in the cuprates \cite{pairing-ansatz,Wilson-01} and 
ultracold gasses \cite{Levin-10}. The essential effects
are here provided by temperature dependent parameter 
$\gamma(T)$ related to scattering caused by the preformed 
pairs and responsible for filling-in the low energy 
states (instead of closing the energy gap as in  
classical superconductors). Its role is hence similar 
to $Z(T)$ of the local solution (\ref{local}). 
%and the perturbative treatment (\ref{interpol}). 
Another parameter $\Sigma_0$ merely controls the line 
broadening so we simply take it as a structureless 
constant.
% \cite{Levin-10}.

For specific numerical computations we used $\Sigma_0\!=
\!\Delta$, assuming $\Delta\!=\!\mbox{\rm const}$. Such
assumption seems reasonable for temperature region exceeding
$T_c$ as long as the binding energy of preformed pairs 
stays constant \cite{Magierski-09} (this constraint
can be modified if necessary). We obtained fairly good
fitting of the experimental data \cite{Jin-10} using
the following empirical temperature dependence 
\begin{eqnarray}
\gamma(T) = 4k_{B}T \; Z(T) .
\label{gamma_T}
\end{eqnarray}
At low temperatures $\gamma(T)$ is  predominantly governed 
by the exponential decay of $Z(T)$, whereas for higher 
temperatures acquires the linear relation $\lim_{t\rightarrow 
\infty} \gamma(T) \propto T$ suggested by various studies 
\cite{Senthil-09,pairing-ansatz}. To establish 
correspondence with the temperature scale we have imposed 
the ratio $2\Delta /k_{B}T_{c}\!=\!4$ realistic for 
the cuprate superconductors and hopefully valid for 
ultracold superfluids near the unitarity. Variation 
of $\gamma(T)$ is illustrated in figure \ref{gammas}.

The {\bf k}-resolved RF measurements provide information 
on the occupied part of the single particle excitation 
spectrum $A({\bf k},\omega)\!=\!-\frac{1}{\pi} \mbox{Im} 
\left[ \omega \!-\! (\varepsilon_{\bf k}\!-\!\mu) \!-\! 
\Sigma({\bf k},\omega) \right]^{-1}$. In figure \ref{kF} 
we show $\omega$-dependence of $A({\bf k},\omega)$ at 
the Fermi momentum ${\bf k}_F$. With constant pairing 
energy $\Delta$ we obtained the gaped spectrum (at 
low temperatures) which gradually evolved into the singly 
peaked structure for $T\! \geq\!1.5 T_c$.

In figure \ref{spectral_fun} we show the momentum 
dependence of the spectral function $A({\bf k},\omega)$  
at temperatures $T/T_{c}=0.76$, $1.24$, $1.47$ and 
$2.06$. Two-peak shape of the EDC-curve $A({\bf 
k}_F,\omega)$ versus $\omega$ is always accompanied 
by presence of the Bogoliubov quasiparticle branches 
with their characteristic bending-down (for $\omega \!<\! 
0$) and bending-up features (for $\omega \!>\! 0$),  
the latter unfortunately hardly accessible by ARPES 
and $k$-resolved RF measurements. This  is driven 
by the  preexisting pairs above $T_c$ correlated 
over the short- range scales. The Bogoliubov-type 
quasiparticles represent admixtures of the particle 
and hole states. In the underdoped cuprates their 
presence has been manifested indirectly through 
the residual diamagnetic response \cite{torque} 
or the large Nernst coefficient and also directly 
in the single particle spectroscopy 
\cite{ARPES-08,Davis-09}.

\begin{figure}
\epsfxsize=9cm\centerline{\epsffile{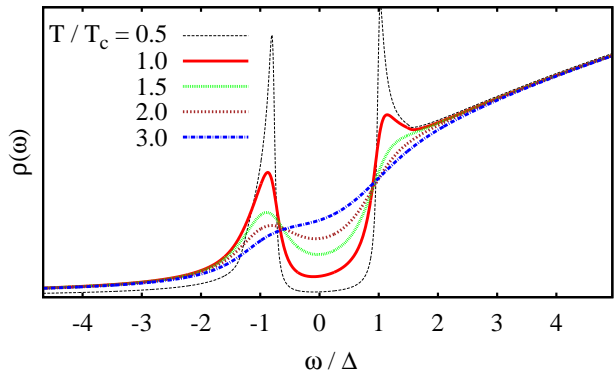}}
\caption{(Color online) The ${\bf k}$-integrated 
density of states for several temperatures as 
indicated.}
\label{ldos}
\end{figure}

We would like to stress that presence of the 
Bogoliubov branches above $T_{c}$ does not go hand
in hand with a suppression of the local density 
of states $\rho(\omega)=\sum_{\bf k}A({\bf k},
\omega)$. This is illustrated in figure \ref{ldos}. 
For temperatures above $1.5T_c$ when we have 
the singly peaked EDC (figure \ref{kF}) and MDC 
(figure \ref{spectral_fun}) the local density of 
states is still clearly depleted around $\omega\sim 0$,
even for temperatures as high as $3T_c$. Such property
might reflect the known discrepancy between large 
values of $T^*$ (signaling the opening of pseudogap)
and the actual estimations of $T^{*}_{sc}$ at which 
the short-range superconducting correlations 
establish \cite{torque}.

\vspace{0.5cm}
\section{Summary}

Transition to the superconducting/superfluid state 
at the unitary limit \cite{Magierski-09} can be 
accompanied by a number of  {\em pre-pairing} 
signatures \cite{Randeria-10} showing up above $T_{c}$.  
%Superconducting correlations are usually manifested 
%indirectly by the collective phenomena. 
Among the known hallmarks of the symmetry broken 
BCS state are the Bogoliubov-type quasiparticles 
representing the mixed particle and hole entities. 
Recent experimental data \cite{Jin-10,ARPES-08} 
clearly indicate that such Bogoliubov quasiparticles 
survive even beyond the superconducting/superfluid 
state. Near the unitary limit this is caused by 
the preformed pairs correlated over short distances 
which strongly affect the single particle spectra 
through the interconversion processes 
(\ref{Andreev}). 

We have examined the impact of preformed pairs 
on the single particle excitation spectrum. In
the instructive solution (\ref{local}) of the local 
Feshbach scattering problem (\ref{Andreev}) we 
have shown how the free fermion states emerge 
out of the Bogoliubov quasiparticles [strictly 
speaking the bonding and antibonding states 
(\ref{A},\ref{B}))] upon increasing temperature. 
Guided by the  perturbative studies \cite{Senthil-09,
pairing-ansatz} we next employed the pairing ansatz 
(\ref{ansatz}) incorporating the local correlations 
and itinerancy of atoms/molecules. Introducing the 
phenomenological scattering rate (\ref{gamma_T})
we were able to reproduce the data of the momentum 
resolved RF spectroscopy obtained for $^{40}$K 
atoms \cite{Jin-10}.

For future studies we suggest use of the 
Andreev tunneling \cite{Daley-08} as a useful and
complementary technique to the ${\bf k}$-resolved 
RF spectroscopy \cite{Jin-10}. If it were feasible for 
the ultracold fermion systems the Andreev reflections 
could unambiguously establish the temperature extent 
of the superconducting/superfluid correlations 
above $T_{c}$.

\vspace{0.3cm}
%{\em Acknowledgements} -
This work is supported by the Ministry of Science 
and Education under the research grant NN202187833.

\end{document}